\documentclass[fleqn,12pt,draft]{article} 
\usepackage[intlimits,reqno]{amsmath}
\usepackage{cite}
\def\oo{\infty}
\def\Z#1{\zeta(#1)}
\def\A#1{a_{#1}}
\def\B#1{b_{#1}}
\def\D#1{d_{#1}}
\def\Re{\mathrm{Re}}
\def\Im{\mathrm{Im}}
\def\parn{\par\noindent}
\def\app{{\left(\frac{\alpha}{\pi}\right)}}
\def\SYS{{\tt SYS}}
\def\Li{{\mathrm{Li}}}
\def\cl#1#2{{\mathrm{Cl}_{#1}}\left({#2}\right)}
\def\Cl#1{{\mathrm{Cl}_{#1}\left(\frac{\pi}{3}\right)}}
\def\Cldd{{\mathrm{Cl^2_2}\left(\frac{\pi}{3}\right)}}
\def\cat#1{{\mathrm{Cl}_{#1}\left(\frac{\pi}{2}\right)}}
\def\catdd{{\mathrm{Cl}^2_{2}\left(\frac{\pi}{2}\right)}}
\def\ZTD{\zeta^2(3)}
\def\rha#1{\Re H_{#1}\left(e^{i\frac{\pi}{3}}\right)}
\def\rhb#1{\Re H_{#1}\left(e^{i\frac{2\pi}{3}}\right)}
\def\iha#1{\Im H_{#1}\left(e^{i\frac{\pi}{3}}\right)}
\def\ihb#1{\Im H_{#1}\left(e^{i\frac{2\pi}{3}}\right)}
\def\rhe#1{\Re H_{#1}\left(e^{i\frac{\pi}{2}}\right)}
\def\ihe#1{\Im H_{#1}\left(e^{i\frac{\pi}{2}}\right)}
\def\rs{\sqrt{s}}
\def\btell{B_3}
\def\ctell{C_3}
\def\intaxoneones#1{f_1(#1)}
\def\intaxonetwos#1{f_2(#1)}
\def\NCL{{\it Nuovo Cimento Lett.}}
\def\NC{{\it Nuovo Cimento }}
\def\NuPh{{\it Nucl. Phys. }}
\def\PL{{\it Phys. Lett. }}
\def\PR{{\it Phys. Rev. }}
\def\PRL{{\it Phys. Rev. Lett. }}
\def\ZP{{\it Z. Phys. }}
\def\IJMP{{\it Int. J. Mod. Phys. }}
\def\aql#1{a_e^{(#1)}}
\def\dq#1{d^Dq_{#1}}
\def\Dq#1{[d^Dq_{#1}]}
\def\e{\epsilon}
\def\Gammae{\Gamma_\e}
\def\Fdu#1{{}_2F_1\left(\begin{smallmatrix}
{{ \frac{1}{2}\;\frac{1}{2} }}\\
{{1 }}\end{smallmatrix}; #1\right)}
\def\Fcub#1{{}_2F_1\left(\begin{smallmatrix}
{{ \frac{1}{3}\;\frac{2}{3} }}\\
{{1 }}\end{smallmatrix}; #1\right)}
\def\Ecub#1{{}_2F_1\left(\begin{smallmatrix}
{{ \frac{1}{3}\;-\frac{1}{3} }}\\
{{1 }}\end{smallmatrix}; #1\right)}
\def\Ftd#1{{}_3F_2\left(\begin{smallmatrix}
{{ \frac{1}{2}\;\frac{1}{2}\;\frac{1}{2} }}\\
{{1\;1 }}\end{smallmatrix}; #1\right)}
\def\Fqt#1{{}_4F_3\left(\begin{smallmatrix}
{{ \frac{1}{2}\;\frac{1}{2}\;\frac{1}{2}\;\frac{1}{2} }}\\
{{1\;1\;1 }}\end{smallmatrix}; #1\right)}
\def\Fqq#1{{}_2F_1^2\left(\begin{smallmatrix}
{{ \frac{1}{4}\;\frac{1}{4} }}\\
{{1 }}\end{smallmatrix}; #1\right)}
\def\TITLE{High-precision calculation of the 4-loop contribution
to the electron $g$-$2$ in QED}
\def\IDENTIFY{\par (Stefano Laporta, \TITLE)}
\def\CAPTIONFIGA{The 4-loop self-mass diagrams.}
\def\CAPTIONFIGB{Examples of vertex diagrams belonging to the 25 gauge-invariant
sets. The number indicates the gauge-invariant set to which the diagram belongs.
In the case of the sets 1-16, 24,25, the other diagrams of each set can be obtained
by permuting separately the vertices on the left and right side of the
main electron line, and considering also the mirror images of the diagrams; 
in the sets containing diagrams with vacuum polarization
insertions, one must also move the vacuum polarization insertion to each internal photon
line. In the sets containing light-light diagrams, one must also
consider the permutations of the vertices of the electron loop.}
\def\CAPTIONFIGC{Minimal set of master integrals which contain all the elliptic constants. 
The double dot in $(a')$ means that denominator is
raised to the power three. $(f,f',f'')$ and $(g,g',g'')$ have numerators
respectively equal to $(1,p.k,(p.k)^2)$.} 
\def\CAPTIONTABLEA{First 1100 digits of $\aql{4}$.}
\def\CAPTIONTABLEB{Contribution to $\aql{4}$ of the 25 gauge-invariant sets of Fig.2.}
\def\CAPTIONTABLEC{Numerical values of the constants of Eq.\ref{Tall} and Eq.\ref{UU}.}
\def\eqref#1{Eq.(\ref{#1})}
\def\eqrefb#1#2{Eqs.(\ref{#1})-(\ref{#2})}
\def\itref#1{(\ref{#1})}
\def\itrefb#1#2{(\ref{#1})-(\ref{#2})}
\def\Figuraboh 
{ 
\begin{figure} 
\begin{center} 
 
 \end{center} 
 \caption{\CAPTIONFIGA}
 \label{self104}
 \end{figure} 
 }            
\def\Figuragau 
{ 
\begin{figure} 
\begin{center} 
 \end{center} 
\caption{\CAPTIONFIGB}
\label{figuragau}
 \end{figure} 
 }            
\def\Figuraunkmas 
{ 
\begin{figure} 
\begin{center} 
 \end{center} 
\caption{\CAPTIONFIGC}
 \label{unkmas}
 \end{figure} 
 }            
\newcommand\mytoday{\number\day\space \ifcase\month\or
  January\or February\or March\or April\or May\or June\or
    July\or August\or September\or October\or November\or December\fi
      \space\number\year}

\begin{document}
\title{\vspace{1cm} \TITLE }
\author{Stefano Laporta\thanks{{E-mail: \tt stefano.laporta{\char"40}bo.infn.it}} \\ 
\hfil \\ {\small \it Dipartimento di Fisica, Universit\`a di Bologna, }
\hfil \\ {\small \it Istituto Nazionale Fisica Nucleare, Sezione di Bologna, }
 \hfil \\ {\small \it Via Irnerio 46, I-40126 Bologna, Italy} 
 }
\date{}
\maketitle
\vspace{-7.5cm} \hspace{12.5cm} {\mytoday} \vspace{+7.5cm}
\vspace{1cm}
\begin{abstract}
I have evaluated up to 1100 digits of precision the
contribution of the 891 4-loop Feynman diagrams contributing 
to the electron $g$-$2$ in QED. The total mass-independent 4-loop contribution is 
$$ a_e =
-1.912245764926445574152647167439830054060873390658725345{\ldots} 
\app^4 \ .
$$
I have fit a semi-analytical expression to the numerical value. The
expression contains harmonic polylogarithms of argument
$e^{\frac{i\pi}{3}}$, 
$e^{\frac{2i\pi}{3}}$,
$e^{\frac{i\pi}{2}}$, one-dimensional integrals of products of
complete elliptic integrals and six finite parts of master integrals, evaluated up to 4800 digits.
\end{abstract}
\vspace{1.5cm}
  
\pagenumbering{roman}
\setcounter{page}{0}
\vfill\eject 
\pagenumbering{arabic}
\setcounter{page}{1}

I have evaluated up to 1100 digits of precision the mass-independent contribution to the 
electron $g$-$2$ anomaly of all the 891 diagrams in 4-loop QED,
thus finalizing a twenty-year effort
\cite{Laporta:2001dd,Laporta:2000dc,Laporta:2001rc,Laporta:2003jz,Laporta:2003xa,Laporta:2008sx,Laporta:2009cna}
begun after the completion
of the calculation of 3-loop QED contribution \cite{Laporta:1996mq}.

Having extracted the power of the fine structure constant $\alpha$ 
\begin{equation}
a_e(\textrm{4-loop})=\aql{4}\app^4\ , 
\end{equation}
the first digits of the result are 
\begin{equation}\label{ae4loop}
\aql{4}=-1.912245764926445574152647167439830054060873390658725345171329848{\ldots}\
. 
\end{equation}
The full-precision result is shown in table \ref{tab:a1100}.
The result \itref{ae4loop} is in excellent agreement ($0.9\sigma$)
with the numerical value
\begin{equation}
\aql{4}(\textrm{Ref.\cite{Aoyama:2014sxa}})=-1.91298(84)\ ,
\end{equation}
latest result of a really impressive pluridecennial effort\cite{Kinoshita:1979ej,Kinoshita:1979ei,Kinoshita:1981wx,Kinoshita:1981ww,Kinoshita:1981wm,Kinoshita:2005zr,Aoyama:2007dv,Aoyama:2007mn,Aoyama:2012wj,Aoyama:2014sxa}.

By using the best numerical value of
$a_e(\textrm{5-loop})=7.795(336) \app^5$(Ref.\cite{Aoyama:2014sxa}), 
 the measurement of the fine structure constant\cite{Bouchendira:2010es}
\begin{equation*}
\alpha^{-1}=137.035\;999\;040 (90)\ ,
\end{equation*}
and the values of  mass-dependent QED, hadronic and electroweak
contributions (see Ref.\cite{Aoyama:2014sxa} and references therein), 
one finds
\begin{equation}\label{aeth}
a_e^{\textrm{th}}=1\;159\;652\;181.664(23)(16)(763) \times 10^{-12}
\ ,
\end{equation}
where the first error comes from $\aql{5}$, the second one
from the hadronic and electroweak corrections, the last one from $\alpha$.
Conversely, using the experimental measurement of $a_e$\cite{Hanneke:2010au}
\begin{equation*}
a_e^{\textrm{exp}}=1\;159\;652\;180.73(0.28) \times 10^{-12}   \ ,
\end{equation*}
one finds 
\begin{equation*}
\alpha^{-1}(a_e) =
 137.035\;999\;1596(27)(18)(331) \ ,
\end{equation*}
where the errors come respectively from  $a_e(\textrm{5-loop})$, hadronic and electroweak corrections, 
and $a_e$.

The 891 vertex diagrams contributing to $\aql{4}$ are not
shown for reasons of space.
They can be obtained by inserting an external photon in each possible electron line 
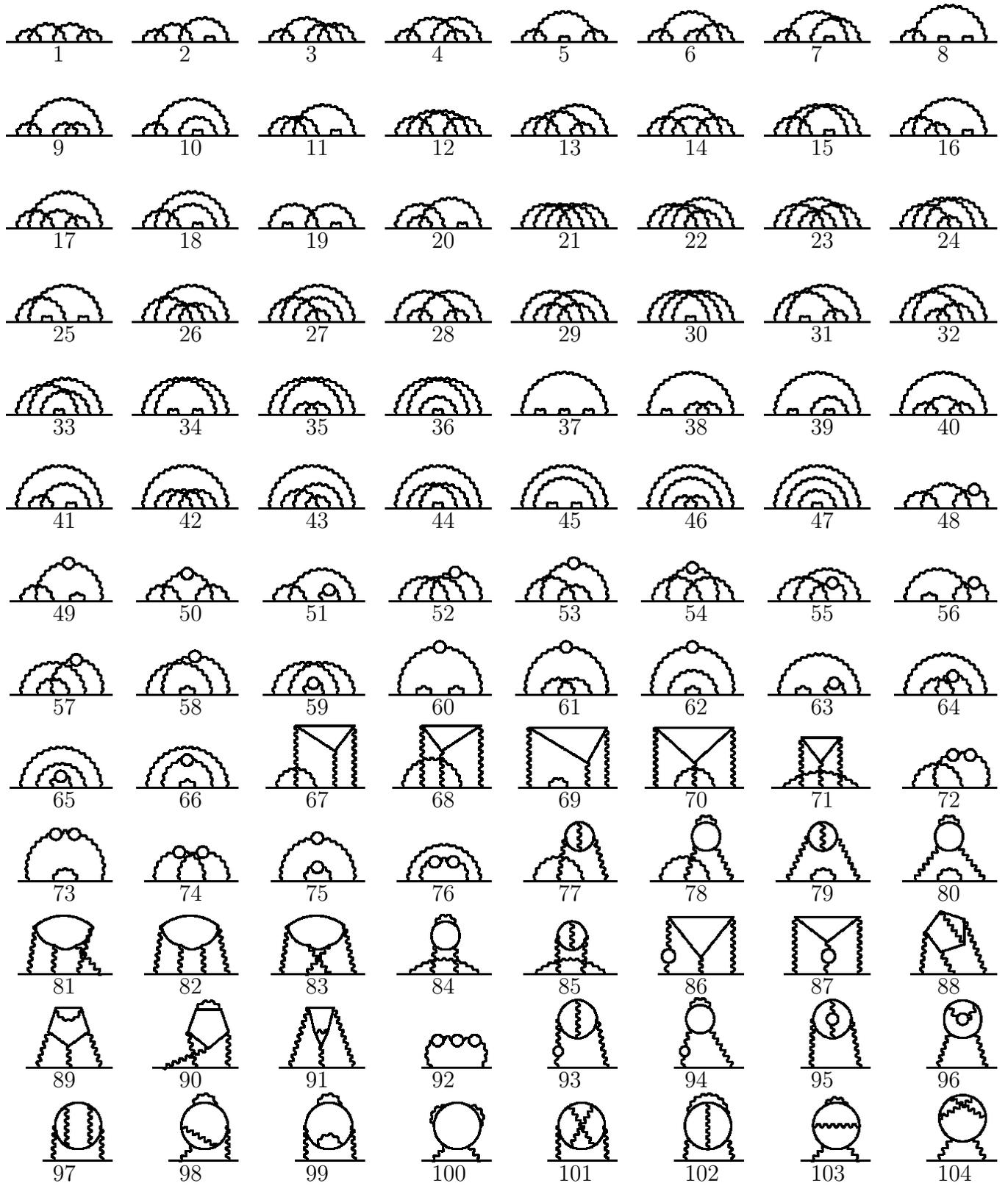
of the 104 4-loop self-mass diagrams shown in Fig.\ref{self104},
excluding the vertex diagrams with closed electron loops with an odd number of
vertices which give null contribution because of the Furry's theorem.
The vertex diagrams can be arranged in 25
gauge-invariant sets (Fig.\ref{figuragau}),
classifying them according to the number of photon corrections on the
same side of the main electron line and the insertions of electron loops (see
Ref.\cite{Cvitanovic:1977dp} for more details on the 3-loop classification).
The numerical contributions of each set, truncated to 40 digits, are
listed in the table \ref{tableset}.
Adding respectively the contributions of diagrams with and without
closed electron loops one finds 
\begin{align} 
 \aql{4}(\textrm{no closed electron loops})&= -2.176866027739540077443259355895893938670 \ ,\\
 \aql{4}(\textrm{closed electron loops only})  &=  \phantom{+}0.264620262813094503290612188456063884609 \ .
\end{align}   
The contributions of the sets 17 and 18, the sum of contributions of
the sets 11 and 12, and the sum of the contributions of the sets 15 and 16 are in
perfect agreement with the analytical results of Ref.\cite{Caffo:1984nm}.

The contributions of all diagrams can be expressed by means of 334
master integrals belonging to 220 topologies.
I have fit analytical expressions to the high-precision numerical
values of all master integrals and diagram contributions by using the PSLQ algorithm\cite{PSLQ,Bailey:1999nv}.
The analytical expression of $\aql{4}$
contains values of harmonic polylogarithms\cite{Remiddi:1999ew}
with argument $1$, $\frac{1}{2}$, $e^{\frac{i\pi}{3}}$ ,
$e^{\frac{2i\pi}{3}}$,  $e^{\frac{i\pi}{2}}$,
a family of one-dimensional integrals of products of
elliptic integrals, and the finite terms of the $\e-$expansions of six master integrals belonging to 
the topologies 81 and 83 of Fig.\ref{self104}. 
Work is still in progress to fit analytically these six unknown
elliptical constants.
The result of the analytical fit can be written as follows:
\begin{align}\label{Tall}
\aql{4}=&T_0+T_2+T_3+T_4+T_5+T_6+T_7
 +\sqrt{3} \left( V_{4a}+V_{6a}\right)  +V_{6b}+V_{7b}
                                        +W_{6b} +W_{7b}
  \cr&
 +\sqrt{3} \left(E_{4a}+E_{5a}+E_{6a}+E_{7a}\right) +E_{6b}+E_{7b} +U\ .
\end{align}
The terms have been arranged in blocks with equal transcendental
weight. The index number is the weight.
The terms containing the ``usual'' transcendental constants are:
\begin{equation} \label{T023}
  T_0+T_2+T_3 =
       \frac{1243127611}{130636800}
       +\frac{30180451}{25920} \Z2
       -\frac{255842141}{2721600} \Z3
       -\frac{8873}{3} \Z2 \ln2 \ ,
\end{equation} 
\begin{equation}\label{T4}
  T_4 =
       \dfrac{6768227}{2160} \Z4
       +\frac{19063}{360} \Z2 \ln^2 2
       +\frac{12097}{90}\left( \A4 +\frac{1}{24} \ln^4 2 \right) \ ,
\end{equation} 
\begin{align}\label{T5} 
  T_5 =&
       -\frac{2862857}{6480} \Z5
       -\frac{12720907}{64800} \Z3 \Z2 
       -\frac{221581}{2160}  \Z4 \ln2
       \nonumber
  \\\cr&
        +\frac{9656}{27} \left(\A5 +\frac{1}{12} \Z2 \ln^3 2
	-\frac{1}{120} \ln^5 2\right) \ ,
\end{align} 
\begin{align}\label{T6} 
  T_6 =&
        \frac{191490607}{46656} \Z6
       +\frac{10358551}{43200} \ZTD
       -\frac{40136}{27} \A6
       +\frac{26404}{27} \B6
       \nonumber
  \\\cr&
       -\frac{700706}{675} \A4 \Z2 
       -\frac{26404}{27}   \A5 \ln2
       +\frac{26404}{27}   \Z5 \ln2
       -\frac{63749}{50}   \Z3 \Z2 \ln2 
  \\\cr&
       -\frac{40723}{135}\Z4 \ln^2 2
       +\frac{13202}{81} \Z3 \ln^3 2
       -\frac{253201}{2700} \Z2 \ln^4 2
       +\frac{7657}{1620} \ln^6 2 \ ,
       \nonumber
\end{align} 
\begin{align}\label{T7} 
  T_7 =&
        \frac{2895304273}{435456} \Z7
       +\frac{670276309}{193536}  \Z4 \Z3
       +\frac{85933}{63} \A4 \Z3
       +\frac{7121162687}{967680} \Z5 \Z2
       \nonumber
  \\\cr&
       -\frac{142793}{18} \A5 \Z2
       -\frac{195848}{21} \A7
       +\frac{195848}{63} \B7
       -\frac{116506}{189} \D7
       \nonumber
  \\\cr&
       -\frac{4136495}{384} \Z6 \ln2
       -\frac{1053568}{189} \A6 \ln2
       +\frac{233012}{189}  \B6 \ln2
       +\frac{407771}{432}  \ZTD \ln2
  \\\cr&
       -\frac{8937}{2} \A4 \Z2 \ln2
       +\frac{833683}{3024} \Z5 \ln^2 2
       -\frac{3995099}{6048} \Z3 \Z2 \ln^2 2
       -\frac{233012}{189} \A5 \ln^2 2
       \nonumber
  \\\cr&
       +\frac{1705273}{1512} \Z4 \ln^3 2
       +\frac{602303}{4536}  \Z3 \ln^4 2
       -\frac{1650461}{11340}\Z2 \ln^5 2
       +\frac{52177}{15876} \ln^7 2 \ .
       \nonumber
\end{align} 
The terms containing harmonic polylogarithms of $e^{\frac{i\pi}{3}}$, $e^{\frac{2i\pi}{3}}$:
\begin{align} 
   V_{4a}=\label{V4a}
       -\frac{14101}{480} \Cl4
       -\frac{169703}{1440} \Z2 \Cl2\ ,
\end{align}       
\begin{align}\label{V6a} 
  V_{6a}=&
        \frac{494}{27} \iha{0,0,0,1,-1,-1}
       +\frac{494}{27} \ihb{0,0,0,1,-1,1}
       +\frac{494}{27} \ihb{0,0,0,1,1,-1}
       \nonumber
  \\\cr&
       +      {19} \ihb{0,0,1,0,1,1}
       +\frac{437}{12} \ihb{0,0,0,1,1,1}
       +\frac{29812}{297} \Cl6
       \nonumber
  \\\cr&
       +\frac{4940}{81} \A4 \Cl2
       -\frac{520847}{69984} \Z5 \pi
       -\frac{129251}{81} \Z4 \Cl2
       \nonumber
  \\\cr&
       -\frac{892}{15}  \ihb{0,1,1,-1} \Z2 
       -\frac{1784}{45} \iha{0,1,1,-1} \Z2
       +\frac{1729}{54} \Z3 \iha{0,1,-1}
  \\\cr&
       +\frac{1729}{36} \Z3 \ihb{0,1,1}
       +\frac{837190}{729} \Cl4 \Z2 
       +\frac{25937}{4860} \Z3 \Z2 \pi
       \nonumber
  \\\cr&
       -\frac{223}{243} \Z4 \pi \ln2
       +\frac{892}{9} \iha{0,1,-1} \Z2  \ln2
       +\frac{446}{3} \ihb{0,1,1}  \Z2  \ln2 
       \nonumber
  \\\cr&
       -\frac{7925}{81}  \Cl2 \Z2 \ln^2 2
       +\frac{1235}{486} \Cl2 \ln^4 2\ ,
       \nonumber
\end{align}       
\begin{align}\label{V6b} 
   V_{6b}=
        \frac{13487}{60} \rha{0,0,0,1,0,1}
       +\frac{13487}{60}  \Cl4 \Cl2
       +\frac{136781}{360}  \Cldd \Z2 \ ,
\end{align}       
\begin{align}\label{V7b} 
   V_{7b}&=
        \frac{651}{4} \rha{0,0,0,1,0,1,-1}
       +     {651} \rha{0,0,0,0,1,1,-1}
       -\frac{17577}{32} \rhb{0,0,1,0,0,1,1}
       \nonumber
  \\\cr&
       -\frac{87885}{64} \rhb{0,0,0,1,0,1,1}
       -\frac{17577}{8} \rhb{0,0,0,0,1,1,1}
       +\frac{651}{4} \Cl4 \iha{0,1,-1} 
       \nonumber
  \\\cr&
       +\frac{1953}{8}  \Cl4 \ihb{0,1,1}
       +\frac{31465}{176}\Cl6 \pi 
       +\frac{211}{4} \rha{0,1,0,1,-1} \Z2 
  \\\cr&
       +\frac{211}{2}   \rha{0,0,1,1,-1} \Z2 
       +\frac{1899}{16} \rhb{0,1,0,1,1} \Z2 
       +\frac{1899}{8}  \rhb{0,0,1,1,1} \Z2 
       \nonumber
  \\\cr&
       +\frac{211}{4} \iha{0,1,-1} \Cl2 \Z2  
       +\frac{633}{8} \ihb{0,1,1}  \Cl2 \Z2  \ .
       \nonumber
\end{align} 
The terms containing harmonic polylogarithms of $e^{\frac{i\pi}{2}}$:
\begin{align}\label{W6b}
   W_{6b} =&
       -\frac{28276}{25} \Z2 \cat2^2\ ,
   \\\cr
   W_{7b} =&104\biggl(
        {4} \rhe{0,1,0,1,1} \Z2 
       +{4} \ihe{0,1,1} \cat2   \Z2 
       -{2}  \cat4 \Z2 \pi
  \\\cr&
       + \catdd \Z2  \ln2\
       \biggr)\ .
       \nonumber
\end{align} 
The terms containing elliptic constants:
\begin{equation}\label{E4a} 
   E_{4a} =\pi\left(
       -\frac{28458503}{691200}    \btell
       +\frac{250077961}{18662400} \ctell
       \right)\ ,
\end{equation} 
\begin{equation}\label{E5I} 
  E_{5a} = \frac{483913}{77760} \pi  \intaxonetwos{0,0,1}
       \ ,
\end{equation} 
\begin{align}\label{E6a} 
   E_{6a}=&\pi\biggl(
        \frac{4715}{1944} \ln2\; \intaxonetwos{0,0,1}
       +\frac{270433}{10935}   \intaxonetwos{0,2,0}
       -\frac{188147}{4860}    \intaxonetwos{0,1,1}
       +\frac{188147}{12960}   \intaxonetwos{0,0,2}
       \biggr)\ ,
\end{align} 
\begin{equation}\label{E6b} 
   E_{6b}=
       -\frac{4715}{1458} \Z2 \intaxoneones{0,0,1}\ ,
\end{equation} 
\begin{align}\label{E7I}
   E_{7a}=&\pi\biggl(       
        \frac{826595}{248832} \Z2 \intaxonetwos{0,0,1}
       -\frac{5525}{432} \ln2\;     \intaxonetwos{0,0,2}
       +\frac{5525}{162} \ln2\;     \intaxonetwos{0,1,1}
       \nonumber
  \\\cr&
       -\frac{5525}{243} \ln2\;     \intaxonetwos{0,2,0}
       +\frac{526015}{248832}     \intaxonetwos{0,0,3}
       -\frac{4675}{768}          \intaxonetwos{0,1,2}
       +\frac{1805965}{248832}    \intaxonetwos{0,2,1}
       \nonumber
  \\\cr&
       -\frac{3710675}{1119744}   \intaxonetwos{0,3,0}
       -\frac{75145}{124416}      \intaxonetwos{1,0,2}
       -\frac{213635}{124416}     \intaxonetwos{1,1,1}
       +\frac{168455}{62208}      \intaxonetwos{1,2,0}
  \\\cr&
       +\frac{75145}{248832}      \intaxonetwos{2,0,1}
       +\frac{69245}{124416}      \intaxonetwos{2,1,0}
       \biggr)\ ,
       \nonumber
\end{align} 
\begin{align}\label{E7b}
   E_{7b}=&\Z2\left(
        \frac{2541575}{82944} \intaxoneones{0,0,2}
       -\frac{556445}{6912}   \intaxoneones{0,1,1}
       +\frac{54515}{972}     \intaxoneones{0,2,0}
       -\frac{75145}{20736}   \intaxoneones{1,0,1}
       \right)\ .
\end{align} 
The term containing the  $\e^0$ coefficients of the $\e-$expansion of
six master integrals (see $f,f',f'',g,g',g''$ of Fig.\ref{unkmas}):
\begin{equation}
   U =\label{UU}
       -\frac{541}{300} C_{81a}
       -\frac{629}{60}  C_{81b}
       +\frac{49}{3}    C_{81c}
       -\frac{327}{160} C_{83a}
       +\frac{49}{36}   C_{83b}
       +\frac{37}{6}    C_{83c}\ .
\end{equation}
The numerical values of \eqrefb{T023}{UU} are listed in Table \ref{tableter}.
In the above expressions  $\zeta(n)=\sum_{i=1}^\oo i^{-n}$,
 $a_n=\sum_{i=1}^\oo 2^{-i}\;i^{-n}$, 
 $b_6=H_{0,0,0,0,1,1}\left(\frac{1}{2}\right)$, 
 $b_7=H_{0,0,0,0,0,1,1}\left(\frac{1}{2}\right)$, 
 $d_7=H_{0,0,0,0,1,-1,-1}(1)$, 
 $\cl{n}{\theta}=\Im \Li_n (e^{i\theta})$.
$H_{i_1,i_2,{\ldots} }(x)$ are the harmonic polylogarithms.
The integrals $f_j$ are defined as follows:
\begin{eqnarray}\label{fdef}
f_1(i,j,k)&=&\int_1^9 ds \; D_1^2(s) \left(s-\frac{9}{5}\right)
\ln^i\left(9-s\right)
\ln^j\left(s-1\right)
\ln^k\left(s\right)
\ ,
\nonumber
\\\cr
f_2(i,j,k)&=&\int_1^9 ds\; D_1(s) \Re\left(\sqrt{3} D_2(s)\right)  \left(s-\frac{9}{5}\right)
\ln^i\left(9-s\right)
\ln^j\left(s-1\right)
\ln^k\left(s\right)
\ ,
\\\cr
D_1(s)&=&\frac{2}{\sqrt{(\rs+3)(\rs-1)^3}}K\left(\frac{(\rs-3)(\rs+1)^3}{(\rs+3)(\rs-1)^3}\right)\
,
\nonumber
\\\cr
D_2(s)&=&\frac{2}{\sqrt{(\rs+3)(\rs-1)^3}}K\left(1-\frac{(\rs-3)(\rs+1)^3}{(\rs+3)(\rs-1)^3}\right)\
;
\nonumber
\end{eqnarray}
$K(x)$ is the complete elliptic integral of the first kind.
Note that $D_1(s)=2J_2^{(1,9)}(s)$, with $J_2^{(1,9)}$ defined in Eq.(A.12) of Ref.\cite{Laporta:2004rb}.
The integrals $f_1(0,0,0)$ and $f_2(0,0,0)$ were studied in
Ref.\cite{Laporta:2008sx}.
The constants $A_3$, $B_3$ and $C_3$, defined in Ref.\cite{Laporta:2008sx},
admit the hypergeometric representations:
\begin{align}\label{cdef}
A_3=\int_0^1 dx &\dfrac{K_c(x) K_c(\textup{1-$x$})}{\sqrt{1-x}}= 
\dfrac{2\pi^\frac{3}{2}}{3}\left(
\dfrac{{\Gamma^2(\frac{7}{6})}{\Gamma(\frac{1}{3})}}{{\Gamma^2(\frac{2}{3})}{\Gamma(\frac{5}{6})}}
{}_4F_3\left(\begin{smallmatrix}
{{\frac{1}{6}\;\frac{1}{3}\;\frac{1}{3}\;\frac{1}{2}}}\\
{{\frac{5}{6}\;\frac{5}{6}\;\frac{2}{3}}}\end{smallmatrix}; 1\right)
-
\dfrac{{\Gamma^2(\frac{5}{6})}{\Gamma(-\frac{1}{3})}}{{\Gamma^2(\frac{1}{3})}{\Gamma(\frac{1}{6})}}
{}_4F_3\left(\begin{smallmatrix}
{{\frac{1}{2}\;\frac{2}{3}\;\frac{2}{3}\;\frac{5}{6}}}\\
{{\frac{7}{6}\;\frac{7}{6}\;\frac{4}{3}}}\end{smallmatrix}; 1\right)
\right)\!\:,\\\cr
B_3=\int_0^1 dx &\dfrac{K_c^2(x)}{\sqrt{1-x}} =
\dfrac{4\pi^\frac{3}{2}}{3}\left(
\dfrac{{\Gamma^2(\frac{7}{6})}{\Gamma(\frac{1}{3})}}{{\Gamma^2(\frac{2}{3})}{\Gamma(\frac{5}{6})}}
{}_4F_3\left(\begin{smallmatrix}
{{\frac{1}{6}\;\frac{1}{3}\;\frac{1}{3}\;\frac{1}{2}}}\\
{{\frac{5}{6}\;\frac{5}{6}\;\frac{2}{3}}}\end{smallmatrix}; 1\right)
+
\dfrac{{\Gamma^2(\frac{5}{6})}{\Gamma(-\frac{1}{3})}}{{\Gamma^2(\frac{1}{3})}{\Gamma(\frac{1}{6})}}
{}_4F_3\left(\begin{smallmatrix}
{{\frac{1}{2}\;\frac{2}{3}\;\frac{2}{3}\;\frac{5}{6}}}\\
{{\frac{7}{6}\;\frac{7}{6}\;\frac{4}{3}}}\end{smallmatrix}; 1\right)
\right) \ , \\\cr
C_3=\int_0^1 dx &\dfrac{E_c^2(x)}{\sqrt{1-x}}=
\frac{486\pi^2}{1925}\;
{}_7F_6\left(\begin{smallmatrix}
    {{\frac{7}{4}\;-\frac{1}{3}\;\frac{1}{3}\;\frac{2}{3}\;\frac{4}{3}\;\frac{3}{2}\;\frac{3}{2}}}\\
    {{\frac{3}{4}\;1\;\frac{7}{6}\;\frac{11}{6}\;\frac{13}{6}\;\frac{17}{6}}}
    \end{smallmatrix}; 1\right) \ ,
\\\cr
K_c(x)&=\frac{2\pi}{\sqrt{27}} \Fcub{x} \ , \qquad E_c(x)=\frac{2\pi}{\sqrt{27}} \Ecub{x} \ . 
\end{align}
$A_3$  appears only in the coefficients of the $\e$-expansion of master
integrals, and cancels out in the diagram contributions.
Fig.\ref{unkmas} shows the fundamental elliptic master integrals which contains irreducible
combinations of $B_3$, $C_3$ and $f_m(i,j,k)$.


The analytical fits of $V_{6b}$, $V_{6a}$, $V_{7b}$, $V_{7i}$  and the
master integrals involved needed PSLQ runs with
basis of $\sim 500$ elements calculated with $9600$ digits of precision. 
The multi-pair parallel version\cite{Bailey:1999nv} of the PSLQ algorithm 
has been essential to work out these difficult analytical fits in reasonable times.

The method used for the computation of the master integrals with
precisions up to 9600 digits is 
essentially based on the difference equation method\cite{Laporta:2001dd,Laporta:2000dc}
and the differential equation method\cite{Kotikov:1990kg,Remiddi:1997ny,Gehrmann:1999as}.
This method and the procedures used for the extraction of $g$-$2$ 
contribution, renormalization, reduction to master integrals, generation and
numerical solution of systems of difference and differential equations,
(all based on upgrades of the program {\tt SYS} of
Ref.\cite{Laporta:2001dd}) will be thoroughly described elsewhere.
\setcounter{secnumdepth}{0} 
\section{Acknowledgments}
The author wants to thank Antonino Zichichi and Luca Trentadue for
having provided support to this work, Ettore Remiddi for continuous support and encouragement.

The main part of the calculations was performed on the cluster ZBOX2 of
the Institute for Theoretical Physics of Zurich and on the Schr\"odinger
supercomputer of the University of Zurich.
The author is deeply indebted to Thomas Gehrmann for having allowed
him to use these facilities.

Some parts of the calculations were done on computers
of the Department of Physics and INFN in Bologna.
The author thanks Michele Caffo, Franco Martelli, Sandro Turrini and
Vincenzo Vagnoni for providing suitable desktop computers in Bologna. 

\vfill\eject 
\pagenumbering{roman}
\setcounter{page}{1}
\begin{table}
\begin{center}
\begin{tt}
-1.9122457649264455741526471674398300540608733906587253451713298480060\\
3844398065170614276089270000363158375584153314732700563785149128545391\\
9028043270502738223043455789570455627293099412966997602777822115784720\\
3390641519081665270979708674381150121551479722743221642734319279759586\\
0740500578373849607018743283140248380251922494607422985589304635061404\\
9225266343109442400023563568812806206454940132249775943004292888367617\\
4889923691518087808698970526357853375377696411702453619601349757449436\\
1268486175162606832387186747303831505962741878015305514879400536977798\\
3694642786843269184311758895811597435669504330483490736134265864995311\\
6387811743475385423488364085584441882237217456706871041823307430517443\\
0557394596117155085896114899526126606124699407311840392747234002346496\\
9531735482584817998224097373710773657404645135211230912425281111372153\\
0215445372101481112115984897088422327987972048420144512282845151658523\\
6561786594592600991733031721302865467212345340500349104700728924487200\\
6160442613254490690004319151982300474881814943110384953782994062967586\\
7875385249781946989793132162197975750676701142904897962085050785592{\ldots}
\end{tt}
\end{center}
\caption{\CAPTIONTABLEA}
\label{tab:a1100}
\end{table}
\Figuraboh
\phantom{.}
\vfill\eject 
\qquad
\Figuragau
\phantom{.}
\vfill\eject 
\qquad
\Figuraunkmas
\vfill\eject 
\begin{table}
\small
\begin{center}
\begin{tabular}{rrrr}
\hline
   1  & - 1.971075616835818943645699655337264406980  \\
   2  & - 0.142487379799872157235945291684857370994  \\
   3  & - 0.621921063535072522104091223479317643540  \\
   4  & \phantom{+} 1.086698394475818687601961404690600972373  \\
   5  & - 1.040542410012582012539438620994249955094  \\
   6  & \phantom{+} 0.512462047967986870479954030909194465565  \\
   7  & \phantom{+} 0.690448347591261501528101600354802517732  \\
   8  & - 0.056336090170533315910959439910250595939  \\
   9  & \phantom{+} 0.409217028479188586590553833614638435425  \\
   10 & \phantom{+} 0.374357934811899949081953855414943578759  \\
   11 & - 0.091305840068696773426479566945788826481  \\
   12 & \phantom{+} 0.017853686549808578110691748056565649168  \\
   13 & - 0.034179376078562729210191880996726218580  \\
   14 & \phantom{+} 0.006504148381814640990365761897425802288  \\
   15 & - 0.572471862194781916152750849945181037311  \\
   16 & \phantom{+} 0.151989599685819639625280516106513042070  \\
   17 & \phantom{+} 0.000876865858889990697913748939713726165  \\
   18 & \phantom{+} 0.015325282902013380844497471345160318673  \\
   19 & \phantom{+} 0.011130913987517388830956500920570148123  \\
   20 & \phantom{+} 0.049513202559526235110472234651204851710  \\
   21 & - 1.138822876459974505563154431181111707424  \\
   22 & \phantom{+} 0.598842072031421820464649513201747727836  \\
   23 & \phantom{+} 0.822284485811034346719894048799598422606  \\
   24 & - 0.872657392077131517978401982381415610384  \\
   25 & - 0.117949868787420797062780493486346339829  \\
\hline
\end{tabular}
\end{center}
\caption{\CAPTIONTABLEB}
\label{tableset}
\end{table}
\vfill\eject 
\begin{table}
\small
\begin{center}
\begin{tabular}{rrrr}
\hline
  $T_0$   &      9.515906781243876151283558690966098373\\
  $T_2$   &   1915.310648253997777888130354499120276542\\
  $T_3$   &  -3485.275086789599708317057778907752410742\\
  $T_4$   &   3504.090225594272699233395974800847330934\\
  $T_5$   &   -725.569913602974274507866288615667084989\\
  $T_6$   &   1381.628304197738147258897402093908402776\\
  $T_7$   &   1692.786400388934476652564199811210670453\\
$V_{4a}$  &   -223.655742930151691157141102901111870825\\
$V_{6a}$  &     14.029138087062071859189974573196626739\\
$V_{6b}$  &    842.150210099809624937684343426149287354\\
$V_{7b}$  &    463.951882993580804359224932846794527895\\
$W_{6b}$  &  -1560.934864680405790411777238139658336036\\
$W_{7b}$  &  -1024.004093725178841133583200254534168436\\
$E_{4a}$  &   -856.605968292200108497784694038000040595\\
$E_{5a}$  &    601.136193120690233763409588135510244820\\
$E_{6a}$  &   -457.790342894702531083496436277945999328\\
$E_{6b}$  &    -89.049936952630079330356943951138211140\\
$E_{7a}$  &    548.453177743013238987339022298522918205\\
$E_{7b}$  &  -2145.946406417837479874008380333397996999\\
$U$       &  - 132.027597619729495491707871522090745221\\
$C_{81a}$ &    116.694585791186600526332510987652818034\\
$C_{81b}$ &    - 8.748320323814631572671010051472284815\\
$C_{81c}$ &    - 0.236085277120339887503638687666535683\\
$C_{83a}$ &      2.771191986145520146810618363218497216\\
$C_{83b}$ &    - 0.807847353263827557176395243854200179\\
$C_{83c}$ &    - 0.434702618543809180642530601495074086\\
\hline
\end{tabular}
\end{center}
\caption{\CAPTIONTABLEC}
\label{tableter}
\end{table}
\end{document}